\documentclass[12pt]{article}
\usepackage{amssymb,latexsym}
\def\nc{N_\mathrm{coarsest}}

\def\mf{M_\mathrm{finest}}

\makeatletter 
 \g@addto@macro\@verbatim\small 
\makeatother

\def\ifundefined#1{\expandafter\ifx\csname#1\endcsname\relax}
\def\bye{\end{document}}   
\long\def\new#1\endnew{{\bf #1}}
\long\def\del#1\enddel{}

\del
\def\HS#1 {\hspace*{#1pt}} \def\VS#1 {\vspace*{#1pt}}
\def\BC{\begin{center}}    
\def\EC{\end{center}}             
        
\enddel

\let\0=\over      \let\1=\vec      \def\2{{1\over2}}    \let\3=\ss
\let\4=\underline \let\5=\overline \let\6=\partial      \def\7#1{{#1}\llap{/}}
\def\8#1{{\textstyle{#1}}}         \def\9#1{{\ifmmode{\pmb{#1}}\else\bf#1\fi}}

          \def\({\left(}       \def\){\right)}

\def\eeql#1 {\label{#1}\eeq}        
\def\beq{\begin{equation}}      \def\eeq{\end{equation}}        
\def\bea{\begin{eqnarray}}      \def\eea{\end{eqnarray}}

\let\and=\wedge                   
\let\then=\Rightarrow      \let\iff=\Leftrightarrow
                
\let\bra=\langle        \let\ket=\rangle        \def\<#1\>{\bra #1 \ket}

\def\rel#1 #2{\buildrel #1 \over {#2}}

          \let\d=\delta   
           
   \let\l=\lambda  \let\m=\mu      
                  \let\r=\rho     \let\s=\sigma

             \let\D=\Delta

\def\IR{{\mathbb R}}  \def\IP{{\mathbb P}} 
   
\def\IZ{{\mathbb Z}}

\def\plb#1 #2 {Phys. Lett. {\bf B#1} #2 }
\def\phr#1 #2 {Phys. Rep. {\bf  #1} #2 }        
\def\npb#1 #2 {Nucl. Phys. {\bf B#1} #2 }
\def\aph#1 #2 {Ann. Phys. {\bf #1} #2 }         
\def\jmp#1 #2 {J. Math. Phys. {\bf #1} #2 }
\def\jgp#1 #2 {J. Geom. Phys. {\bf #1} #2 }
\def\prd#1 #2 {Phys. Rev. {\bf D#1} #2 }
\def\prl#1 #2 {Phys. Rev. Lett. {\bf #1} #2 }
\def\rmp#1 #2 {Rev. Mod. Phys.  {\bf #1} #2 }
\def\zpc#1 {Z. Phys. {\bf #1C} }
\def\cmp#1 #2 {Commun. Math. Phys. {\bf #1} #2 }
\def\cqg#1 #2 {Class.Quant.Grav. {\bf #1} #2 }
\def\mpl#1 {Mod. Phys. Lett. {\bf A#1} }
\def\cpc#1 {Computer Phys. Commun. {\bf #1} }   
\def\ijmp#1 {Int. J. Mod. Phys. {\bf A#1} }
\def\ijmpC#1 {Int. J. Mod. Phys. {\bf C#1} }
\def\atmp#1 {Adv. Theor. Math. Phys. {\bf #1} }

\begin{document}
{$ $}
\vskip 15mm
\begin{center}
{\Large\bf How to Classify Reflexive Gorenstein Cones\\[1cm]}
Harald Skarke
\\[3mm]
Institut f\"ur Theoretische Physik, Technische Universit\"at Wien\\
Wiedner Hauptstra\ss e 8--10, A-1040 Wien, AUSTRIA\\[3mm]
{\tt skarke@hep.itp.tuwien.ac.at}
\vfill 
                 {\bf ABSTRACT } 
\end{center}
Two of my collaborations with Max Kreuzer involved classification problems 
related to string vacua. 
In 1992 we found all 10,839 classes of polynomials that lead to 
Landau-Ginzburg models with c=9 (Klemm and Schimmrigk also did this); 
7,555 of them are related to Calabi-Yau hypersurfaces. 
Later we found all 473,800,776 reflexive polytopes in four dimensions; 
these give rise to Calabi-Yau hypersurfaces in toric varieties. 
The missing piece -- toric constructions that need not be hypersurfaces -- 
are the reflexive Gorenstein cones introduced by Batyrev and Borisov. 
I explain what they are, how they define the data for Witten's gauged linear 
sigma model, and how one can modify our classification ideas to apply to them. 
I also present results on the first and possibly most interesting step,
the classification of certain basic weights systems, and discuss limitations 
to a complete classification. 
\vfill
\newpage
\section{Introduction}
\subsection{Landau--Ginzburg models}
I first met Max in 1987, when he was in the last stages of his doctorate 
studies at the Institute for Theoretical Physics of the TU Vienna, and I 
was just starting mine.
We were then both working in quantum field theory, and Max already displayed 
his well known capacity for compressing essential information; in particular,
he had one piece of paper containing all the formulas one would ever need for 
computing certain Feynman diagrams.
Max then went on to postdoctoral positions at Hanover and Santa Barbara.
When he came back to Vienna, I had finished my thesis and was looking
for something new to work on.
Max, who had started to collaborate with Rolf Schimmrigk in Santa Barbara,
invited me to join a project related to string compactifications.
Since neither of us knew much string theory at the time (but both of us had 
math diplomas), our strategy was to isolate a mathematical nucleus from 
an important topic, namely orbifolds of N=2 superconformal field theories 
({\it SCFTs}) of Landau--Ginzburg type that can be used for string 
compactifications.
Such a model requires a potential that is a quasihomogeneous function
$f({\phi_i})$ of the fields
with an isolated singularity at the origin:
\beq f({\l^{q_i}\phi_i})= \l f({\phi_i}),~~~~~~
\frac{\6 f}{\6\phi_i}=0~\forall i~~~\then~~~\phi_i=0~\forall i.\eeql{nondeg}
This gives rise to a superconformal field theory whose anomaly $c$
is determined by the positive rational numbers $q_i$ (the `weights')
via $c/3 = \sum_{i=1}^N(1-2q_i)$.
In a first paper with Rolf \cite{KSS} we considered symmetries of some known 
models of that type, but then we turned to the classification problem.
This had been treated for very simple cases in a book by Arnold et al. 
\cite{AGV}.
For realistic string compactifications one requires $c=9$, i.e. 
$\sum_{i=1}^N(1-2q_i)=3$.
Well known examples include
\beq f=\phi_1^5+\ldots+\phi_5^5 ~~~\hbox{ and }~~~ f=\phi_1^3+\ldots+\phi_9^3. 
\eeq
At that time it was known that the cases $N=4,5$ were related to Calabi--Yau 
manifolds --- the functions could be used to define hypersurfaces in 
weighted projective spaces ---
whereas the cases $N=6,7,8,9$ were not (at least, not directly);
however, the precise form of the relationship between Landau--Ginzburg models 
and Calabi--Yaus was unclear.
Max and I extended the approach of \cite{AGV} and used the resulting
algorithm \cite{cqf} to find 10,838 sets of $q_i$ that fit the criteria.
In the meantime Klemm and Schimmrigk had done a more thorough search of the 
literature which allowed them to attack the problem directly, and while we 
were in the last stages of writing up our results they published theirs 
\cite{KlS} which contained precisely one model more.
By taking another look at some candidates that we had rejected we managed to 
find that model and arrived at the same set of 10,839 \cite{nms}.
One thing that was noticeable from these results was the fact that mirror 
symmetry was not complete within this class of models.

\subsection{The gauged linear sigma model}
The question about the precise relationship between Landau--Ginzburg models 
and Calabi--Yaus was settled quite beautifully by Witten \cite{Wph}.
He introduced a (2,2)--supersymmetric gauged linear sigma model in two 
dimensions
which contained chiral superfields $\Phi_i$ (with component fields $\phi_i$, 
$F_i$) and vector superfields $V_a$ (with components $\s_a$, $D_a$).
The theory's superpotential $W(\Phi)$ is invariant under 
$\Phi_i\to e^{-iQ_{i,a}\l_a}\Phi_i$, and there are real parameters $r_a$ 
coming from the Fayet-Iliopoulos terms.
It turns out that the model determines the $D$- and $F$-terms as
\beq 
D_a = -e_a^2(\sum_iQ_{i,a}|\phi_i|^2-r_a), ~~~F_i =\frac{\6 W}{\6\phi_i}, 
\eeql{DFterms}
and requires minimization of the potential 
\beq 
U = \sum_a\frac{1}{ 2e_a^2}D_a^2 +\sum_i|F_i|^2+2\sum_{i,a}|\s_a|^2|\phi_i|^2Q_{i,a}^2.
\eeql{Upot}
The behaviour of the theory depends crucially on the values of the $r_a$ as
the following classic example demonstrates.\\[2mm]
{\bf Example 1.}
Consider the case of just one vector field and six chiral superfields with
charges $Q_0=-5$, $Q_1=\ldots =Q_5=1$, and a superpotential
$W=\Phi_0P^5(\Phi_1,\ldots,\phi_5)$, where $P^5$ is a polynomial 
of degree 5 that is non-degenerate, i.e. obeys eq. (\ref{nondeg}) . Then
\[ D \sim r+5|\phi_0|^2-\sum_{i=1}^5|\phi_i|^2, ~~~|F_0|^2=|P^5|^2, 
~~~\sum_{i=1}^5|F_i|^2 = |\phi_0|^2\sum_{i=1}^5|\frac{\6 P^5}{\6\phi_i}|^2,\]
and we can distinguish the following cases.\\
$r >> 0$: Then $D^2\to$ min implies $(\phi_1,\ldots,\phi_5)\ne (0,\ldots,0)$, so
by non-degeneracy of $P^5$ we need $\phi_0=0$ to minimize $\sum_{i=1}^5|F_i|^2$ .
The ground state is located at 
\[ \{(\phi_1,\ldots,\phi_5): \sum_{i=1}^5|\phi_i|^2 = \sqrt{r}, 
   P^5(\phi_i)=0\}/U(1),\]
which is just the symplectic quotient description of a quintic hypersurface 
in $\IP^4$, i.e. the standard Calabi--Yau example.\\
$r << 0$: Then $D^2\to$ min requires $\phi_0\ne 0$, implying
$\frac{\6 P^5}{\6\phi_i}=0$ and therefore $\phi_1=\ldots=\phi_5=0$.
The $U(1)$ gauge symmetry may be used to fix $\phi_0=\sqrt{-r/5}$, leaving a 
residual $\IZ_5$ symmetry.
The resulting model is just a $\IZ_5$ orbifold of a Landau--Ginzburg model 
with potential $P^5$, i.e. one of the 10,839 models we had classified.

By introducing more than one gauge field and more than one analogue of $\phi_0$
one can easily build models that correspond to complete intersection 
Calabi--Yaus.

\subsection{Toric constructions}
Around that time Victor Batyrev \cite{Bat} introduced a construction that was 
manifestly mirror symmetric in the following sense.
Given a dual pair of lattices $M\simeq\IZ^d$, $N=\mathrm{Hom}(M,\IZ)$ and 
their real extensions $M_\IR\simeq\IR^d$, $N_\IR\simeq\IR^d$, one defines a
{\it lattice polytope} $\D\subset M_\IR$ as a polytope with vertices in $M$,
and a {\it reflexive polytope} as a lattice polytope $\D\ni 0_M$ whose dual
\beq \D^* = \{y\in N_\IR: \<y,x\>+1\ge 0 ~~\forall x\in\D\} \eeql{dual}
is also a lattice polytope.
To any triangulation of the surface of $\D^*$ one can assign the toric variety 
$\cal V$ that is determined by the corresponding fan, with homogeneous 
coordinates $z_i$ that are in one to one correspondence with the nonzero 
lattice points $y_i$ of $\D^*$.
Then every lattice point $x_j$ of $\D$ determines a monomial 
$M_j=\prod_i z_i^{\<y_i,x_j\>+1}$, and the generic polynomial consisting of
these monomials defines a Calabi--Yau hypersurface $\cal C\subset\cal V$.
The Hodge numbers of $\cal C$ can be computed directly from the structure of
$\D$ and it turns out that the exchange $(M, \D)\leftrightarrow (N, \D^*)$
effects precisely the flip of the Hodge diamond that is associated with mirror
symmetry.

\del
To such a reflexive pair one can assign a class of Calabi--Yau hypersurfaces
by considering the toric variety that is determined by the fan over some 
maximal crepant triangulation of $\D^*$, and the hypersurface in that 
toric variety that is the zero locus of a polynomial in the homogeneous 
coordinates such that every monomial corresponds to a lattice point of $\D$,
with 
\enddel

Borisov \cite{Bor} generalized this construction to complete intersections 
$\cal C$ in toric varieties.
The main idea is to generalize the duality of eq. (\ref{dual}) to sets of
polytopes $\nabla_i\subset N_\IR$, $\D_j\subset M_\IR$ for 
$i, j\in \{1,\ldots, {\rm codim }~ \cal C\}$ via 
\beq \<y,x\>+\d_{ij}\ge 0 ~~\forall ~y\in\nabla_i, ~x\in\D_j; \eeql{nefp}
the fan for $\cal V$ is given by a triangulation of Conv($\{\nabla_i\}$)
which turns out to be reflexive.
At this point it is not clear how this is related to the gauged linear sigma 
model; in particular, fields like $\phi_0$ have no analogues in the toric
coordinates, and Landau--Ginzburg models without Calabi--Yau interpretation
are missing.
This situation changed with two papers by Batyrev and Borisov who
introduced reflexive Gorenstein cones \cite{BB1} and a formula for the
corresponding `stringy Hodge numbers' \cite{BB2} that displays exactly
the type of combinatorial duality required by mirror symmetry.
We postpone precise definitions to the next section and 
just mention here that the data of these cones 
can be used to define gauged linear sigma models.

\subsection{The classification of reflexive polyhedra}
Given all these developments it was clear that the answer to the `missing 
mirror problem' lay in the realm of gauged linear sigma models and toric
geometry.
At that time reflexive polytopes were classified only in dimensions up to two
(there are 16 reflexive polygons), and no algorithm for a classification
in higher dimensions was known.
In the autumn of 1995 Max in I were both in Vienna again and started to work 
on a general algorithm.
We realized that the 
inversion of inclusion relations by duality,
$\D\subseteq\tilde\D\iff\D^*\supseteq\tilde\D^*$,
has the following implication.
If we find a set $S$ of polytopes such that every reflexive polytope contains
at least one member of $S$, then every reflexive polytope must be contained
in one of the duals of the members of $S$.
One can choose $S$ as a set of polytopes $\nabla_{\rm min}$ that 
are {\it minimal} in the sense that $0$ is in the interior of $\nabla_{\rm min}$
but not in the interior of the convex hull of any subset of the set of 
vertices of $\nabla_{\rm min}$.
We proved that every such $\nabla_{\rm min}$ is either a simplex or the convex 
hull of lower dimensional simplices in a specific way \cite{crp}.
In two dimensions the only possibilities are triangles and parallelograms;
in 3d examples would include tetrahedra, octahedra, egyptian pyramids, etc.
Any simplex involved in such a construction determines a {\it weight system} 
$\{q_i\}$ via $0=\sum_iq_iV_i$ where the $V_i$ are the vertices of the simplex.
In order to play a role for the classification of reflexive polytopes a 
minimal polytope must satisfy 
$0 \in {\rm int}({\rm conv}(\nabla_{\rm min}^*\cap M))$. 
This condition restricts the admissible weight systems to a finite set;
a procedure for obtaining them in arbitrary dimensions and the results in
up to four dimensions were presented in \cite{wtc}.
Combining these with all possible combinatorial structures of minimal
polytopes led to a complete list of minimal polytopes.
By considering all subpolytopes of polytopes in the dual list of maximal objects
we could find all 4,319 reflexive polytopes in three dimensions \cite{c3d}
and all 473,800,776 in four dimensions \cite{c4d}.
A thorough description of the complete algorithm in its final form can be found
in \cite{ams}.

This project was at the limit of what could be achieved with the computers that
were available to us, 
so we required extremely efficient routines for handling 
lattice polytopes. 
After some polishing these routines were published as the package PALP 
\cite{palp} which is still being updated every now and then. 
An up-to-date manual of the current version can be found in this volume
\cite{pum}.

\subsection{Structure of the paper}
From everything discussed so far it is clear that the piece that is 
missing from our classification results 
is the case of toric
constructions that need not correspond to Calabi--Yau hypersurfaces; in other 
words it is the Batyrev/Borisov construction of reflexive Gorenstein cones.
This is what the rest of this paper will be about.
In the following section some of the essential definitions are given.
In section 3 the classification problem is analysed in the spirit of 
\cite{crp,wtc,ams}.
Section 4 describes the classification of the relevant new weight systems,
and section 5 discusses the further steps that could be taken.

\section{Some definitions}

A {\it Gorenstein cone} $\s$ is a cone in $M_\IR$ with generators 
$V_1,\ldots,V_k\in M$
satisfying $\<V_i,n_\s\>=1$ for some element $n_\s\in N$.
The {\it support} $\D_\s$ of $\s$ is the polytope Conv($\{V_1,\ldots,V_k\}$) in
the hyperplane $\<x,n_\s\>=1$ in $M_\IR$.

A {\it reflexive Gorenstein cone} $\s$ is a Gorenstein cone whose dual 
\[\s^{\vee}=\{y\in N_\IR:\<y,x\>\ge 0 ~~\forall x\in\s\}\]
is also Gorenstein, i.e. there exists an 
$m_\s\in M$ such that $\<m_\s,W_i\>=1$ for all generators $W_i$ of  
$\s^{\vee}$; the integer $r=\<m_\s,n_\s\>$ is called the {\it index} of $\s$.
If $\s$ is reflexive with index $r$ then $r\D_\s$ is a reflexive polytope
\cite{BB1}.

A reflexive Gorenstein cone $\s$ of index $r$ is called {\it split} if 
$M\simeq\IZ^k\oplus \tilde M$ and $\s$ is generated by
$(e_1,\D_1),\ldots,(e_k,\D_k)$ 
where the $e_i$ form a basis of $\IZ^k$ and the $\D_i$ are lattice polytopes 
in $\tilde M_\IR$.
This implies $k\le r$; $\s$ is called {\it completely split} if $k=r$.

If both $\s$ and $\s^\vee$ are completely split (the latter with a basis 
$ \{f_i\}$ for $\IZ^r$ and polytopes $\nabla_i\subset\tilde N_\IR$) it can be 
shown \cite{BN} that one can choose the bases $\{e_i\}$ and $\{f_i\}$ 
dual to each other.
Then the duality of the cones is equivalent to eq. (\ref{nefp})
which is the defining property of a {\it nef-partition} \cite{Bor}.

\del
A {\it nef-partition} is a complete splitting of both $\s$ and $\s^\vee$ 
(the latter with a basis $ \{f_i\}$ for $\IZ^r$ that is 
dual to $ \{e_i\}$ and polytopes $\nabla_i\subset\tilde N_\IR$), such that 
$\< x,y\>\ge -1$ for $x\in\D_i$, $y\in \nabla_i$
and $\< x,y\>\ge 0$ for $x\in\D_i$, $y\in \nabla_j$ with $i\ne j$.

A particular class of reflexive Gorenstein cones of index $r$ can be 
constructed in the following way.
Consider dual lattices $\tilde M$, $\tilde N$ and their underlying real vector 
spaces, and polytopes $\D_i\subset\tilde M$, $\nabla_i\subset\tilde N$, 
$i\in\{1,\ldots,r\}$ such that $\< x,y\>\ge -1$ for $x\in\D_i$, $y\in \nabla_i$
and $\< x,y\>\ge 0$ for $x\in\D_i$, $y\in \nabla_j$ with $i\ne j$.
If we write $\{e_i\}$, $\{f_j\}$ for mutually dual bases of $\IZ^r$ or $\IR^r$,
then $\s=\mathrm{Conv}(\{(e_1,\D_1),\ldots,(e_r,\D_r)\})$,
$\r=\mathrm{Conv}(\{(f_1,\nabla_1),\ldots,(f_r,\nabla_r)\})$ 
form a dual pair of reflexive Gorenstein cones of index $r$ w.r.t. the lattices
$M:=\IZ^r\oplus \tilde M$, $N:=\IZ^r\oplus \tilde N$, respectively.
An isomorphism between any  pair of reflexive Gorenstein $(\s,\s^\vee)$ cones 
and a construction of this type is called a nef-partition of $(\s,\s^\vee)$.

With cones $\s_1\subset M_{1,\IR}$, $\s_2\subset M_{2,\IR}$, we can 
construct the cone $\s=\mathrm{Conv}(\{(\s_1,0),(0,\s_2)\})$ in 
$M_\IR=M_{1,\IR}\oplus M_{2,\IR}$.
This construction is additive in the dimension $d=d_1+d_2$ and the index 
$r=r_1+r_2$, with the dual cone given by 
$\s^\vee=\mathrm{Conv}(\{(\s_1^\vee,0),(0,\s_2^\vee)\})$ in 
$N_\IR = N_{1,\IR}\oplus N_{2,\IR}$.
\enddel

The cartesian product $\s_1\times\s_2\subset M_{1,\IR}\oplus M_{2,\IR}$ of two
reflexive Gorenstein cones is again a reflexive Gorenstein cone, with 
dimension $d=d_1+d_2$, index $r=r_1+r_2$ and dual cone 
$\s_1^\vee\times\s_2^\vee\subset N_{1,\IR}\oplus N_{2,\IR}$.

Given a reflexive pair $(\s,\s^\vee)$ of Gorenstein cones and denoting by 
$\{x_j\}$ ($\{y_i\}$) the set of lattice points in the support of $\s$ 
($\s^\vee$) and by $l_a: \sum_i Q_{i,a}y_i=0$ a basis for the set of linear 
relations among the $y_i$, one can define a gauged linear sigma model by 
introducing 
\begin{itemize}
\item a chiral superfield $\Phi_i$ for every $y_i$,
\item a gauge field $V_a$ for every $l_a$,
\item the charges $Q_{i,a}$ as the coefficients of the $l_a$,
\item a monomial $M_j=\prod_i\Phi_i^{\<y_i,x_j\>}$ for every $x_j$.
\end{itemize}

\section{Analysis of the classification problem}\label{analysis}

Let us fix $n\in N$ and $m\in M$ with $\<m,n\>=r$.
The main ideas of \cite{crp,wtc,ams} can be adapted as follows.
We say that a Gorenstein cone $\s\subset M_\IR$ with $n_\s=n$ has the 
{\it IP}
(for `interior point') property if $m$ is in the interior of $\s$.
This is equivalent to $m/r$ being in the interior of the support $\D_\s$.
With an analogous definition of the IP property for a cone $\r\subset N_\IR$
we call $\r$ {\it minimal} if it has the IP property, but if no cone generated
by a proper subset of the set of generators of $\r$ has it.
The support $\nabla_\r$ of $\r$ is a minimal polytope in the sense of 
\cite{crp},
characterized by the fact that the set $\{V_1,\ldots,V_{d+k-1}\}$ of its 
vertices is the union of $k\ge 1$ subsets (possibly overlapping) such that each 
of them determines a simplex (of lower dimension unless $k=1$) 
with the interior point (here, $n/r$) in its relative interior.
\del
For the classification of reflexive polytopes with $\mathbf 0$ as IP one then 
proceeds to define a weight system as a collection of positive numbers $q_i$
such that $\sum q_i V_i = 0$, where the $V_i$ are the vertices belonging to a
particular simplex.
\enddel
This implies that $n$ lies in the interior of the cone generated by the 
vertices $V_i$ of such a simplex, so there exists a uniquely defined set
of positive rational numbers $q_i$ such that $\sum q_i V_i = n$;
acting with $m$ on this equation we see that $\sum q_i = r$.
We call the $q_i$ the {\it weight system} associated with the simplex; 
if $k>1$ the collection of weight systems is referred to as a 
{\it combined weight system} or {\it CWS}.

In the case $k=1$ where $\r$ itself is simplicial we have an identification
of $\r\subset N_\IR$ with $\IR_{\ge 0}^d\subset \IR^d$ via 
$V_i\leftrightarrow e_i$. 
The corresponding identification of dual spaces implies 
$m \leftrightarrow (1,\ldots,1)$.
Up to now we have not specified the lattice $N$. 
Given $n$ and the generators $V_i$, clearly the coarsest possible 
lattice $\nc$ is the one generated by these vectors, 
corresponding to the lattice in $\IR^d$ generated by 
$e_1,\ldots,e_d$ and ${\mathbf q} = (q_1,\ldots,q_d)$.
The lattice $\mf$ dual to $\nc$ is then determined by the isomorphism
\[ \mf\simeq\{(x_1,\ldots,x_d): x_i\in \IZ, \sum x_i q_i \in \IZ \}. \]
Let us now define $\s(\r)$ as the cone over 
\[\mathrm{Conv}(\r^\vee\cap\{x\in M: \<x,n\> = 1\}),\] 
and $\s(\mathbf q)$ as $\s(\r)$ for the case $ M = \mf$.
We say that $\mathbf q$ has the IP property if $\s(\mathbf q)$ has it, i.e.
if $(1,\ldots,1)$ is interior to the cone over 
$\{(x_1,\ldots,x_d):x_i \in \IZ_{\ge 0}, \sum x_iq_i=1\}$.
This is equivalent to
$(1/r,\ldots,1/r) \in \mathrm{int}(\D_{\mathbf q})$
with
\beq
\D_{\mathbf q}=
\mathrm{Conv}(\{(x_1,\ldots,x_d):x_i \in \IZ_{\ge 0}, \sum x_iq_i=1\}).  
\eeql{ipws}
We note that this does not rely on $r$ being integer, allowing us to
talk about {\it IPWSs} (`IP weight systems') for $(d,r)$ with rational $r$.

The cartesian product of cones has an 
analogue in the fact that if 
${\mathbf q^{(1)}}$, ${\mathbf q^{(2)}}$ are IPWSs for $(d_1,r_1)$
and $(d_2,r_2)$, respectively, then 
$\mathbf q = ({\mathbf q^{(1)}}, {\mathbf q^{(2)}})$ is an IPWS
for $(d_1+d_2,r_1+r_2)$.
Note, however, that generically $\mf(\mathbf q)$ is finer than 
$\mf(\mathbf q^{(1)})\oplus \mf(\mathbf q^{(2)})$.\\[2mm]
{\bf Lemma 1.} Assume that $(q_1,\ldots,q_{d})$ form a $(d,r)$--IPWS.
Then\\
a) every $q_i$ obeys $q_i\le 1$;\\
b) if $q_d=1$ then $(q_1,\ldots,q_{d-1})$ form a $(d-1,r-1)$--IPWS;\\
c) if $q_d=1/2$ then $(q_1,\ldots,q_{d-1})$ form a $(d-1,r-1/2)$--IPWS;\\
d) if $q_{d-1}+q_d=1$ then $(q_1,\ldots,q_{d-2})$ form a $(d-2,r-1)$--IPWS,
and $q_{d-1}=q_d=1/2$ or $q_{d-1}$ and $q_d$ can be written as nonnegative
integer linear combinations of $q_1,\ldots,q_{d-2}$.\\[2mm]
{\bf Proof.} a) If $q_i > 1$ then $x_i=0$ in $\D_{\mathbf q}$, so 
$(1/r,\ldots,1/r)$ is not in the interior.\\
b), c) Here $\D_{\mathbf q}$ is the pyramid over $\D_{(q_1,\ldots,q_{d-1})}$ with 
apex the point $e_d$ or $2e_d$ which has the IP 
property if and only if $\D_{(q_1,\ldots,q_{d-1})}$ has it.\\
d) The case $q_{d-1}=q_d=1/2$ can be reduced to case c), so let us assume 
$q_{d-1}=1-q_d > 1/2$, $q_d < 1/2$.
If we denote by $\l$ the largest integer satisfying $\l q_d\le 1$, 
then $\D_{\mathbf q}$ is the convex hull of
\[\D_1\cup\{e_{d-1}+e_d\}\cup\{e_{d-1}+\D_{q_d}\}
\cup\bigcup_{\mu=1}^\l\{\m e_d+\D_{1-\m q_d}\} \]
where we have written $\D_y$ for 
$\mathrm{Conv}(\{(x_1,\ldots,x_{d-2},0,0):x_i \in \IZ_{\ge 0}, 
x_1q_1+\ldots x_{d-2}q_{d-2}=y\})$.
If $q_d$ could not be written as a nonnegative integer linear combination of
$q_1,\ldots,q_{d-2}$ then $\D_{q_d}$ would be empty and every point of 
$\D_{\mathbf q}$ would satisfy $x_d\ge x_{d-1}$.
But then $(1/r,\ldots,1/r)$ would lie at the boundary of
$\D_{\mathbf q}$, thus violating the IP assumption.
Similarly, if all $\D_{1-\m q_d}$ were empty we would have the same type of 
contradiction via $x_{d-1}\ge x_d$, so at least one of the $\D_{1-\m q_d}$ must 
be non-empty, but then 
$\D_{1-q_d}\supseteq\D_{1-\m q_d}+(\m-1)\D_{q_d}$
implies that $\D_{1-q_d}$ must also be non-empty, hence $1-q_d=q_{d-1}$ is a 
nonnegative linear combination of $q_1,\ldots,q_{d-2}$.
\newline
Finally, let us assume that $(q_1,\ldots,q_{d-2})$ does not form a 
$(d-2,r-1)$--IPWS.
Then there is some hyperplane through $0$ and $(1/(r-1),\ldots,1/(r-1))$ 
such that all of $\D_1$ lies on the same side of it:
$a_1x_1+\ldots +a_{d-2}x_{d-2}\ge 0 $ for all $x\in \D_1$,
with $a_i$ satisfying $a_1+\ldots+a_{d-2}=0$.
As the point $x_1=\ldots=x_d=1/r$ lies in the same hyperplane, the IP property 
can hold for $\D_{\mathbf q}$ only if there is at least one point with 
$a_1x_1+\ldots+a_{d-2}x_{d-2}<0$.
If this point pertains to $\D_{q_d}$, denote by $c$ the maximal value for which 
$a_1x_1+\ldots+a_{d-2}x_{d-2}=-c$.
Then $\D_1\supseteq\D_{1-\m q_d}+\m\D_{q_d}$ implies 
$a_1x_1+\ldots+a_{d-2}x_{d-2}\ge\m c$ for all $x\in \D_{1-\m q_d}$ and inspection
of the components of $\D_{\mathbf q}$ shows that they all obey
\[a_1x_1+\ldots+a_{d-2}x_{d-2}+cx_{d-1}-cx_d\ge 0,\] 
thereby violating the IP condition for $\D_{\mathbf q}$.
Similarly, if one or more of the $\D_{1-\m q_d}$ contain points with 
$a_1x_1+\ldots+a_{d-2}x_{d-2}<0$, we choose $c$ to be the maximal value for which
$a_1x_1+\ldots+a_{d-2}x_{d-2}=-\m c$.
Then $\D_1\supseteq\D_{1-\m q_d}+\m\D_{q_d}$ implies 
$a_1x_1+\ldots+a_{d-2}x_{d-2}\ge c$ for all $x\in \D_{q_d}$ and 
all components of $\D_{\mathbf q}$ obey
\[a_1x_1+\ldots+a_{d-2}x_{d-2}-cx_{d-1}+cx_d\ge 0,\] 
again violating the IP condition for $\D_{\mathbf q}$.
\hfill$\Box$

Note, however: $q_{d-1}+q_d=1$ does not imply that one of these two
repeats one of the other weights as the IPWS $(111126)[8]$ shows
(the notation $(n_1\ldots n_d)[k]$ means $q_i=n_i/k$);
$q_d>1/2$ does not imply $1-q_d\in\{q_1,\ldots,q_{d-1}\}$ as 
demonstrated by the IPWS $(111114)[6]$.

Motivated by the lemma we shall refer to a weight system as {\it basic}
if it contains no weights $q_i \in \{1/2,1\}$ and no $q_i$, $q_j$ with 
$q_i+q_j=1$.
For such a weight system, any $\mathbf x$ satisfying $\sum x_i q_i =1$ must
obey $\sum x_i > 2$.

What happens if $\r$ is not simplicial, i.e. $\nabla_\r$ consists of
more than one simplex?
Then one embeds each of the $k>1$ simplices
$S_i$ into $\IR^{d_i}\subset\IR^{d+k-1}$, where $\IR^{d_i}$ is the subspace 
spanned by the $e_j$ corresponding to the $d_i$ vertices of $S_i$;
the interior points ${\mathbf q}^{(i)}$ of the resulting simplicial cones 
are identified and on gets
\[ N\simeq  (\IZ^{d+k-1}\oplus \IZ {\mathbf q}^{(1)}\cdots
\oplus \IZ {\mathbf q}^{(k)})/\{a_{ij}({\mathbf q}^{(i)}-{\mathbf q}^{(j)}):
a_{ij}\in \IZ\}. \]
On the $M$ lattice side 
one now has $k$ equations of the type $\sum x_iq_i=1$ in $\IZ_{\ge 0}^{d+k-1}$.
In particular, if the simplices all have distinct vertices, one starts
with the cartesian product of cones in $N$ and projects along the differences of
the $n_i$, $i\in \{1, \ldots,k\}$; in $M$ this results in the support $\D_\s$
being the product of the supports $\D_{\s_i}$, $i\in \{1, \ldots,k\}$.

Given these preparations the following algorithm for the classification of 
reflexive Gorenstein cones in dimension $d$ with index $r$ emerges.
\begin{enumerate}
\item Find all basic IPWSs for $d'\in \{0,1,\ldots,d\}$, 
$r'\in \{0,1/2,1,\ldots,r\}$ with $r-r'\le d-d'$.
\item Extend the results of the first step by weights 1, 1/2 and $(q, 1-q)$
to get all IPWSs with index $r$ and dimension $d'\le d$.
\item Determine all possible structures of minimal polytopes in dimension
$d-1$.
\item Combine the last two steps to get all $d$-dimensional minimal cones.
\item Determine all subcones on all sublattices of $\mf$.
\end{enumerate}

\section{Classification of basic weight systems}
The classification of basic IPWSs relies on the algorithm of \cite{wtc}.
In order to find all $\mathbf q$'s satisfying 
$(1/r,\ldots,1/r) \in \mathrm{int}(\D_{\mathbf q})$ with $\D_{\mathbf q}$
determined by eq.~(\ref{ipws}) one uses the fact
that $\mathbf q$ is determined
by a set of linearly independent $\mathbf x$'s satisfying $\sum x_iq_i=1$.
The classification proceeds by successively choosing such ${\mathbf x}^{(i)}$, 
starting with ${\mathbf x}^{(0)}=(1/r,\ldots,1/r)$ and continuing with lattice
points ${\mathbf x}^{(1)},\ldots {\mathbf x}^{(k)}$.
Every choice of a new $\mathbf x$ restricts the set of allowed $\mathbf q$'s.
Given ${\mathbf x}^{(0)},\ldots, {\mathbf x}^{(k)}$ one can choose any
$\tilde {\mathbf q}$ compatible with them and check whether it has the IP 
property.
A further $\mathbf q \ne\tilde {\mathbf q}$ can have the IP property only 
if $k+1<d$ and $\D_{\mathbf q}$ contains points on both sides of the 
hyperplane $\sum x_i\tilde q_i=1$.
In particular, such a $\mathbf q$ must be compatible with one of the finitely 
many lattice points obeying $x_i\ge 0$ for all $i$ and $\sum x_i\tilde q_i<1$.
For every choice of ${\mathbf x}^{(k+1)}$ among these one should then continue
in the same way.\\[2mm]
{\bf Example 2.} 
$d=2$, $r=1/2$: ${\mathbf x}^{(0)}=(2,2)$ is compatible with
$\tilde {\mathbf q} = (1/4,1/4)$, which has the IP property.
Any further $\mathbf q$ must allow at least one integer point with 
$x_1+x_2<4$.
Up to permutation of coordinates the only possiblities are 
${\mathbf x}^{(1)}=(3,0)$ which leads to ${\mathbf q} = (1/6,1/3)$,
and ${\mathbf x}^{(1)}=(2,1)$ which does not result in a positive weight system.
\\[2mm]
{\bf Lemma 2.} 
If $d=3r$ there is precisely one basic IPWS $(1/3,\ldots,1/3)$, and
for $d<3r$ there is no basic IPWS.\\[2mm]
{\bf Proof. }
Let us assume $d\le 3r$. 
The point $(1/r,\ldots,1/r)$ is compatible with 
$\tilde {\mathbf q} = (r/d,\ldots,r/d)$, which has the IP property if $d=3r$.
Any other $\mathbf q$ must admit at least one point $\mathbf x$ such that
$1 > \sum x_i \tilde {\mathbf q}_i \ge (\sum x_i)/3$, i.e. $\sum x_i \le 2$,
which is not consistent with a basic IPWS.
\hfill$\Box$
\\[2mm]
The cases covered neither by example 2 nor by lemma 2
require the use of a computer.
PALP \cite{palp} contains an implementation of the algorithm of \cite{wtc}
that works reasonably well for $r\le 1$ and $d\le 5$. 
In order to get a program that is fast enough even for the case $r=3$, $d=8$
the corresponding routines had to be rewritten completely.
In particular, the present implementation takes into account some of the 
symmetry coming from permutations of the coordinates.
At every choice of ${\mathbf x}^{(k)}$ in the recursive construction the 
program computes the vertices of the $(d-k-1)$--dimensional polytope in 
$\mathbf q$--space that is determined by $q_i\ge 0$ and 
$\sum_i x_i^{(j)}q_i=1$ for $j\in \{0,\ldots,k\}$. 
This can be done efficiently by using the $(d-k)$--dimensional polytope of
the previous recursive step.
$\tilde {\mathbf q}$ is chosen as the average of the vertices of the 
$\mathbf q$--space polytope.

This program was used to determine all basic IPWSs for $r\le 3$ and
$d\le 9$. 
The complete lists can be found at the website \cite{KScy}.
The results are summarized in table \ref{nIPWS}
\begin{table}[ht]
\begin{center}
{\begin{tabular}{|l|rrrr|} 
\hline 
~~r{\Large$\backslash$}$d_{\rm CY}$&~~~~~~~~0&~~~~~~~~1&~~~~~~~~2&~~~~~~~~3\\ 
\hline
            ~~0&1&0&0&0\\
          1/2&0&2&48&97,036\\
            ~~1&0&1&47&86,990\\
          3/2&0&0&28&168,107\\
            ~~2&0&0&1&34,256\\
          5/2&0&0&0&6,066\\
            ~~3&0&0&0&1\\
\hline
\end{tabular}}
\end{center}
\caption{Numbers of basic IPWSs for given values of $r$ vs. $d_{\rm CY}=d-2r$}
\label{nIPWS}
\end{table}
which shows the numbers of basic IPWSs for given index $r$ and $d-2r$.
Following \cite{BN} we call the latter `Calabi--Yau dimension' $d_{\rm CY}$;
in the case of a complete splitting of the cone it is indeed the dimension of 
a complete intersection Calabi--Yau variety defined by the corresponding 
nef-partition, and for any cone leading to a sensible superconformal field 
theory it is $c/3$ where $c$ is the conformal anomaly.

The first entry is the empty IPWS for $d=r=0$ which is required as a starting 
point for the construction of IPWSs containing only weights $1/2$ or $1$.

For $d_{\rm CY}=1$ there are the three basic weight systems $(1/4,1/4)$,
$(1/6,1/3)$ and $(1/3,1/3,1/3)$ from example 2 and lemma 2.

For $d_{\rm CY}=2$ there are 48 basic IPWSs with $r=1/2$ and 47 with $r=1$.
Together they determine precisely the well known 95 weight systems for 
weighted $\IP^4$'s that have K3 hypersurfaces \cite{reid,fl89};
as weight systems for reflexive polytopes they were determined in \cite{wtc}.
In addition there are 28 basic IPWSs with $r=3/2$ as well as the IPWS
$(1/3,\ldots,1/3)$ for $r=2$.
These 29 additional basic IPWSs are again identical with the ones relevant
to Landau--Ginzburg type SCFTs as determined in \cite{nms};
each of them gives rise to a reflexive Gorenstein cone.

Finally, for $d_{\rm CY}=3$ there are the 184,026 weight systems with $r\le 1$ 
relevant to Calabi--Yau hypersurfaces in toric varieties \cite{wtc}, 
which contain the 7,555 weight systems relevant to weighted projective spaces
\cite{KlS,nms} as a small subset.
In addition there are $168,107+34,256+6,066+1=208,430$ IPWSs with $r>1$
which are new (except for $3,284$ Landau--Ginzburg weights \cite{KlS,nms}).
These weight systems can be the starting points for constructing codimension
2 and 3 Calabi--Yau threefolds in toric varieties as well as N=2 SCFTs with 
$c=9$.
While each of the 184,026 weight systems with $r\le 1$ determines a reflexive 
polytope (hence a reflexive Gorenstein cone) as shown already in \cite{wtc},
among the Gorenstein cones determined by IPWSs with $r>1$ only 
$112,817+18,962+1,321+1=133,101$ of $208,430$ are reflexive; 
nevertheless the others are relevant to the classification because they may 
contain reflexive subcones.
For the reflexive cases the `stringy Hodge numbers' of \cite{BB2} as computed
by PALP 2.1 \cite{pum} are also listed at the website \cite{KScy}.
The pairs of Hodge numbers all seem to be in the range that is well known 
from the earlier classifications.

\section{Further steps of the algorithm}
We shall  now illustrate further steps of the algorithm presented at the end 
of section \ref{analysis} for some of the smallest $(d,r)$--pairs.
The case of $d_{\rm CY}=1$ corresponds to 
$(d,r)\in\{(3,1), (5,2), (7,3),\ldots \}$.

\subsection{$d=3,~r=1$}
\begin{enumerate}
\item According to the previous section, the relevant basic IPWSs are\\
$d'=0,~r'=0$: $()$; \\
$d'=2,~r'=1/2$: $(1/4,~1/4),~ (1/6,~1/3)$; \\
$d'=3,~r'=1$: $(1/3,~1/3,~1/3)$.
\item These give rise to the $r=1$ IPWSs\\
$d'=2$: $(1/2,~1/2)$;\\
$d'=3$: $(1/3,~1/3,~1/3),~ (1/4,~1/4,~1/2),~  (1/6,~1/3,~1/2)$.
\item A 2--dimensional minimal polytope is a triangle or a rhomboid \cite{crp}.
\item A minimal cone is determined by one of the weight systems \\
(1/3, 1/3, 1/3), (1/4, 1/4, 1/2),  (1/6, 1/3, 1/2) or the CWS \\
(1/2, 1/2, 0, 0; 0, 0, 1/2, 1/2).
\item All reflexive subcones correspond to all reflexive subpolytopes of
the corresponding support polytopes (3 triangles and a square); these are
the well known 16 reflexive polygons.
\end{enumerate}

\subsection{$d=5,~r=2$}
\begin{enumerate}
\item The relevant basic IPWSs are\\
$d'=0,~r'=0$: $()$; \\
$d'=2,~r'=1/2$: $(1/4,~1/4),~ (1/6,~1/3)$; \\
$d'=3,~r'=1/2$: 48 basic IPWSs (cf. table \ref{nIPWS});\\
$d'=3,~r'=1$: $(1/3,~1/3,~1/3)$;\\
$d'=4,~r'=1$: 47 basic IPWSs (cf. table \ref{nIPWS}).
\item These give rise to the $r=2$ IPWSs\\
$d'=2$: $(1,~1)$;\\
$d'=3$: $(1/2,~1/2,~1)$;\\
$d'=4$: $(1/2,~1/2~,~1/2,~1/2)$,\\
${}$~~~~~~~~~~$(1/3,~1/3,~1/3,~1)$, $(1/4,~1/4,~1/2,~1)$,
$(1/6,~1/3,~1/2,~1)$;\\
$d'=5$: $(1/4,~1/4,~1/2,~1/2,~1/2)$, $(1/6,~1/3,~1/2,~1/2,~1/2)$,\\
${}$~~~~~~~~~~$(1/3,~1/3,~1/3,~1/2,~1/2)$,\\ 
${}$~~~~~~~~~~$(1/4,~1/4,~1/4,~1/2,~3/4)$, $(1/6,~1/6,~1/3,~1/2,~5/6)$,\\
${}$~~~~~~~~~~$(1/6,~1/3,~1/3,~1/2,~2/3)$, $(1/3,~1/3,~1/3,~1/3,~2/3)$,\\
${}$~~~~~~~~~~48 IPWSs of the type $(q_1,~q_2,~q_3,~1/2,~1)$,\\
${}$~~~~~~~~~~47 IPWSs of the type $(q_1,~q_2,~q_3,~q_4,~1)$.
\item The 4--dimensional minimal polytopes were classified in\cite{crp}.
\item 5. These steps would require the use of a computer and have not yet 
been performed.
\end{enumerate}

\subsection{Other cases}
The next case with $d_{\rm CY}=1$ is $d=7$, $r=3$.
Here already the first step of the algorithm involves the 184,026 weight 
systems that were used in the classification of reflexive polytopes in 
four dimensions, as well as as the 28 basic IPWSs for $d'=5$, $r'=3/2$.
In addition it requires an analysis of the possible structures of minimal 
polytopes in dimensions up to 6.
This should not be too hard, but one should be aware of the fact that a
description of a minimal polytope in terms of IP simplices need not be unique,
as pointed out already in \cite{crp}.

From what we have seen it is clear that for any fixed value of $d_{\rm CY}$
the complete classification problem gets harder for rising $r$.
In particular the lists for $d_{\rm CY}=3$ contain weight systems of the 
type $(q_1,\ldots,q_6,1)$ for $r=2$ and of the type $(q_1,\ldots,q_7,1,1)$ 
for $r=3$.
In the cases where classifications have been completed it turns out
that there are more weight systems for reflexive ($d-1$)--polytopes 
than there are reflexive $d$--polytopes, 
so while it is conceivable that $(d=5,r=1/2)$
and $(d=6,r=1)$ might be within the range of present computer power, 
$(d=7,r=1)$ is definitely impossible.

However, one would not expect all reflexive Gorenstein cones to lead to
sensible SCFTs.
For example, consider a cone $\sigma$ whose support is a height one pyramid,
which is equivalent to 
$\sigma = \sigma_b\times \sigma_1$ where $\sigma_b$ is the cone over the base 
of the pyramid and $\sigma_1$ is the unique one dimensional cone;
this case leads to trivial $E_{\rm string}$ \cite{BN}.
Now $\sigma({\mathbf q})$ with ${\mathbf q}=(\tilde{\mathbf q},1)$ is
of this type, and any of its subcones with the IP property is also of
this type because 
all lattice points are in the base or the apex; 
hence the apex of the pyramid cannot be dropped without 
violating the IP property.
Therefore one can omit 
cones defined by single weight systems containing a weight of 1 from the 
list of cones serving as starting points for step (5) of the
classification procedure.
This implies that in addition to the basic weight systems of table 
\ref{nIPWS} only $(d=5,r=1/2)$ and $(d=6,r=1)$ are required for a 
classification of relevant CWS for $d_{\rm CY}\le 3$, $r\le 3$.
More generally one might use the fact that $E_{\rm string}$ is multiplicative
under taking cartesian products of cones \cite{NiSch}; the case
above is a special case of this since $E_{\rm string}=0$ for the one dimensional
cone (actually $E_{\rm string}=0$ whenever $d_{\rm CY}< 0$ \cite{NiSch}).

A further reduction of the number of relevant (C)WS may come from the following 
consideration related to the gauged linear sigma model.
If the superpotential contains quadratic terms then its derivatives $F_i$
(cf. eq. \ref{DFterms}) have linear terms that can be used to 
eliminate (`integrate out' in physicists' language) fields by 
replacing them by the expressions determined by $F_i=0$. 
In this way one can argue for the following simplifications:
a support polytope that is 
a height 2 pyramid over a height 2 pyramid can be reduced to the base, 
implying that a weight system $({\mathbf q},1/2,1/2)$ is equivalent to 
just $({\mathbf q})$;
the product of two height one pyramids can be reduced to 
the product of the bases, implying the equivalence 
$({\mathbf q},1,{\mathbf 0},0;{\mathbf 0},0,\tilde{\mathbf q},1)\sim 
({\mathbf q},{\mathbf 0};{\mathbf 0},\tilde{\mathbf q})$ of CWS;
a weight system $(\tilde{\mathbf q},q,1-q)$ should be equivalent to 
$(\tilde{\mathbf q})$.
While these considerations certainly need to be put on a firmer footing, 
they seem to be confirmed `experimentally' as the following lines of
PALP output (version 2.1 is required, see \cite{pum}) indicate.
\begin{verbatim}
4 1 1 1 1 0 0  2 0 0 0 0 1 1 M:105 8 N:7 6 H:2,86 [-168]
4 1 1 1 1 4 0 0 0  2 0 0 0 0 0 1 1 2 M:144 15 N:10 8 H:2 86 [-168]
3 1 1 1 0 0 0  3 0 0 0 1 1 1 M:100 9 N:7 6 H:2,83 [-162]
3 1 1 1 3 0 0 0 0  3 0 0 0 0 1 1 1 3 M:121 16 N:10 8 H:2 83 [-162]
6 1 1 1 1 2 3 3 M:181 7 N:7 7 H:1 103 [-204]
5 1 1 1 1 1 1 4 M:258 12 N:8 8 H:1 101 [-200]
\end{verbatim}
However, one should not draw the conclusion that only basic weight systems are 
relevant: for example, the CWS $(1, 1, 0, 0, 0, 0; 0, 0, 1/2, 1/2, 1/2, 1/2)$ 
corresponds to the perfectly sensible case of two quadrics in $\IP^3$.

\del
Conjecture: quadratic terms can be integrated out\\
pyramid $\times$ pyramid $\to$ base $\times$ base\\
4 1 1 1 1 0 0  2 0 0 0 0 1 1 M:105 8 N:7 6 H:2,86 [-168]\\
4 1 1 1 1 4 0 0 0  2 0 0 0 0 0 1 1 2 M:144 15 N:10 8 H:2 86 [-168]\\
3 1 1 1 0 0 0  3 0 0 0 1 1 1 M:100 9 N:7 6 H:2,83 [-162]\\
3 1 1 1 3 0 0 0 0  3 0 0 0 0 1 1 1 3 M:121 16 N:10 8 H:2 83 [-162]\\
Height 2 pyramid over height 2 pyramid $\to$ base\\
10 2 2 2 2 2 5 5 M:129 7 N:9 7 H:2 202 [-400] h0=2\\
6 1 1 1 1 2 3 3 M:181 7 N:7 7 H:1 103 [-204]\\
$(\tilde{\mathbf q},q,1-q)\to \tilde{\mathbf q}$ \\
5 1 1 1 1 1 1 4 M:258 12 N:8 8 H:1 101 [-200]\\
BUT: (1, 1, 0, 0, 0, 0; 0, 0, 1/2, 1/2, 1/2, 1/2) perfectly sensible:
Two quadrics in $\IP^3$!;\\
1 1 1 0 0 0 0 0 0  1 0 0 1 1 0 0 0 0  1 0 0 0 0 1 1 0 0  1 0 0 0 0 0 0 1 1 M:16 16 N:8 8 H:[0]\\
etc.
Todos:
extension with $q\ge 1/2$--weights,
weights for $d=6$, $r=1$ (also important for 4-folds!)
\enddel

Finally let us discuss what can be done in the future.
Extending the basic weight systems with $q\ge 1/2$--weights is completely 
straightforward but only interesting once we also combine several weight
systems into CWS, which should not be too hard, either.
The classification of $(d=5,r=1/2)$ and $(d=6,r=1)$ basic weight systems
probably is the most interesting step that may still be achieved,
in particular since these same weight systems also give rise to
Calabi--Yau fourfolds.
In principle this could be done with the existing algorithm. 
In practice it is very unlikely that it would produce results within a 
reasonable computation time.
One would probably need to work very hard on further elimination of 
redundancies,
on parallelizing the computation and on obtaining the necessary computer power.
This would require someone with great skills in 
understanding the problem, programming, and organizing resources;
in other words, someone like Max Kreuzer.

\del
{\it Acknowledgements:} 
We would like to thank 
for useful discussions. 
\enddel
        



\begin{thebibliography}{11}

\def\I#1{{\it #1}}      \addtolength{\itemsep}{-4.5pt}  \small \vspace{-3mm}

\ifundefined{draftmode} \def\.#1 #2\>{\bibitem{#1}#2}           
\else        \def\BP{\begin{picture}} \def\EP{\end{picture}}  
  \def\LLab#1{\BP(0,0)\unitlength=1mm\put(-12,.5){\makebox(0,0)[cr]{\small #1
        \rlap{$_{_{\makeatletter\csname TRef#1\endcsname\makeatother}}$}}}\EP}
       \def\.#1 #2\>{\bibitem{#1}\LLab{#1}#2}  \fi

\.KSS   M. Kreuzer, R. Schimmrigk and H. Skarke,
        \I{Abelian Landau-Ginzburg Orbifolds and Mirror Symmetry,}
        \npb 472 (1992) 61, hep-th/9112047.   \>

\.AGV   V.I.Arnold, S.M.Gusein-Zade and A.N.Varchenko,
        \I{Singularities of Differentiable Maps,}
        Vol. I, Birkh\"auser 1985.   \>

\.cqf   M. Kreuzer, H. Skarke, \I{On the Classification of Quasihomogeneous 
        Functions,} \cmp 150 (1992) 137, hep-th/9202039.        \>

\.KlS   A. Klemm, R. Schimmrigk, \I{Landau--Ginzburg String Vacua,}
        \npb 411 (1994) 559, hep-th/9204060.    \>

\.nms   M.Kreuzer, H.Skarke, \I{
        No mirror symmetry in Landau-Ginzburg spectra!,} 
        Nucl. Phys. B388 (1992) 113,  hep-th/9205004.   \>

\.Wph   E. Witten, \I{Phases of N=2 theories in two dimensions,} 
        \npb 403 (1993) 159, hep-th/9301042. \>

\.Bat   V.V. Batyrev, \I{Dual Polyhedra and Mirror Symmetry for Calabi--Yau
        Hypersurfaces in Toric Varieties,}
        J. Alg. Geom. {\bf 3} (1994) 493,  alg-geom/9310003.  \>

\.Bor   Lev Borisov, \I{Towards the Mirror Symmetry for Calabi-Yau Complete 
        intersections in Gorenstein Toric Fano Varieties,} alg-geom/9310001. \>

\.BB1   V.V. Batyrev, L.A. Borisov, \I{Dual Cones and Mirror Symmetry for 
        Generalized Calabi-Yau Manifolds,} Mirror symmetry II 
        (eds. B. Greene, S. T. Yau) 71-86, alg-geom/9402002.   \>

\.BB2   V.V. Batyrev, L.A. Borisov,
        \I{Mirror Duality and string-theoretic Hodge numbers,} 
        Invent. Meth. {\bf 126} (1996) 183, alg-geom/9509009.   \>

\.crp   M. Kreuzer, H. Skarke, \I{On the Classification of Reflexive 
        Polyhedra,} \cmp 185 (1997) 495, hep-th/9512204.        \>

\.wtc   H. Skarke, \I{Weight Systems for Toric Calabi--Yau Varieties and
        Reflexivity of Newton Polyhedra,}
        Mod. Phys. Lett. {\bf A11} (1996) 1637, alg-geom/9603007.   \>

\.c3d   M. Kreuzer, H. Skarke,  \I{Classification of Reflexive Polyhedra 
        in Three Dimensions,} Adv. Theor. Math. Phys. {\bf 2} (1998) 847, 
        hep-th/9805190.  \>

\.c4d   M. Kreuzer, H. Skarke, \I{Complete Classification of
        Reflexive Polyhedra in Four Dimensions,} 
        Adv. Theor. Math. Phys. {\bf 4} (2000) no. 6, hep-th/0002240.  \>

\.ams   M. Kreuzer, H. Skarke, \I{Reflexive polyhedra, weights and
        toric Calabi-Yau fibrations,}
        Rev. Math. Phys. {\bf 14} (2002) 343, math.AG/0001106.   \>

\.palp  M. Kreuzer, H. Skarke, \I{PALP: A Package for Analyzing Lattice 
        Polytopes with Applications to Toric Geometry,}
        Comput.Phys.Commun. 157 (2004) 87, math.SC/0204356. \>

\.pum   A. Braun, J. Knapp, E. Scheidegger, H. Skarke and N.--O. Walliser,
        \I{PALP: a User Manual,} to be published in `Strings, Gauge Fields, 
        and the Geometry Behind - The Legacy of Maximilian Kreuzer' 
        (World Scientific).  \>

\.BN    V.V. Batyrev, B. Nill, \I{Combinatorial aspects of mirror symmetry,}
        math/0703456.   \>

\.KScy  M. Kreuzer, H. Skarke,\\
        {\tt http://hep.itp.tuwien.ac.at/$\sim$kreuzer/CY.html.} \>

\.reid  M. Reid, \I{Canonical 3-folds,} Proc. Alg. Geom. Anger 1979,
        Sijthoff and Nordhoff, 273. \>

\.fl89  A. R. Fletcher, \I{Working with complete intersections,} Bonn preprint
        MPI/89--35 (1989).   \>

\.NiSch B. Nill, J. Schepers, \I{Gorenstein polytopes and their stringy
        E-functions,}
        Math. Ann. (to appear), DOI: 10.1007/s00208-012-0792-2,
        arXiv:1005.5158.   \>


\end{thebibliography}
\bye